# First principles calculation of the valence band offsets for β-polytype of A3B6 layered crystals.

Z.A.Jahangirli*, F.M. Hashimzade, D.A. Huseynova, B.H. Mehdiyev, and N.B. Mustafayev.

*Institute of Physics, National Academy of Sciences of Azerbaijan, Baku, Azerbaijan.*

*E-mail: zakircahangirli@yahoo.com

**Abstract**

The valence band offsets (VBO) for the β-type $A^3B^6$ layered compounds depending on the thickness of the crystals have been investigated from the first principles, based on the density functional theory. To simulate the structure of a given thickness the periodic slab model was used. Two adjacent crystal slabs consisting of several layers were separated by a vacuum region of two-layer width. It is shown that at the crystal thickness more than 12 layers, photothreshold practically becomes independent on the thickness of the crystal.

## 1. Introduction

Monochalcogenides GaS, GaS and InSe have a complex layered structure wherein each layer includes four alternating atomic planes X - M - M - X, where M = Ga, In, and X = S, Se. The unit cell of β-polytype these crystals contain two layers. The space group symmetry is P63/mmc ($D_{6h}^4$). All atoms are in the crystallographic position 4 (f): ± (1/3, 2/3, z) and ± (1/3, 2/3, 1/2-z). The lattice parameters and the positions of atoms in the structure in the fractional coordinates are follows: for GaS  a = 3.587 Å, c = 15.492 Å , z(Ga) = 0.1710, z(S) = 0.6016 [1]; for GaSe a = 3.742 Å, c = 15.919 Å, z(Ga) = 0.1710, z(S) = 0.6016 [2]; for InSe a = 4.05 Å, c = 16.93, z(Ga) = 0.1710, z(S) = 0.6016 Å [2] .

As shown in [3, 4], the ultrathin $A^3B^6$ crystals on various substrates can be used as a highly sensitive photodetectors. In addition, the above mentioned chalcogenides were found to be promising materials for the transformation of solar energy.

For determination of the VBO we calculated the photothreshold for each of above mentoined crystalls. The photothreshold is the minimum energy required to remove an electron from the top of the valence band into the vacuum. Knowing the value of the photothreshold one can determine the contact potential difference in the hetero-structures used as solar energy converters. To the best of our knowledge there is only one study [5], where the photo thresholds of monolayer GaS, GaSe, and InSe were calculated from first principles. In this work have been calculated the valence band offsets (VBO) of these $A^3B^6$ crystals with various thicknesses. Results of the photothresholds calculations for the monolayer $A^3B^6$ crystals are also included for comparison with the results of the theoretical calculations of [5].

## 2. Calculation method

To simulate the structure of a given thickness the periodic slab model was used. Our estimation show that the vacuum thickness of about 18 Å is sufficient for interactions between adjacent slabs to be neglected. Actually, we have constructed a super-lattice of one, two or more unit cells of $A^3B^6$ crystal with the vacuum of one unit-cell-thick.

The calculations has been carried out from first principles using the plane wave pseudopotential code ABINIT [6]. Exchange correlation interaction was described in the local density approximation (LDA) according to [7]. The numerical integration over the Brillouin zone was carried out using the Monkhorst-Pack 12×12×1 grid with the (0., 0., 0.5) shift from the origin [8]. For the pseudopotentials we use the norm-conserving Hartwigsen-Goedecker-Hutter pseudopotentials [9]. In the wave function expansion we included the plane waves with the maximal energy up to 1350 eV, which ensures good convergence of the total energy. The atomic positions and structural parameters have been optimized by calculating the Hellmann-Feynman forces. The equilibrium parameter values were determined by minimisation of the total energy with precision of up to 10-6 eV per unit cell. The the module of forces are minimized with the criterion of 10-4 eV/Å.

## 3. Discussion of the results

The analysis of calculated equilibrium positions of atoms (Table 1-3) show that the structural change of slabs, consisting of multiple layers is insignificant. This is not surprising as the interlayer interaction in these crystals is weak, and distant layers have little effect on the geometry of the surroundings of atoms in the particular layer. The tables 1-3 also shows the thickness and vacuum settings for monolayers of [5].

The dependencies of the value of the photothreshold on the thickness of GaS, GaSe, and InSe are shown in Fig.1. The value of the photothreshold correspond to the energy difference between the electrostatic potential in a vacuum region far from the neutral surface of crystal and the Fermi level which coincides with the top of the valence band in this case.

In [5] for the monolayer of the $A^3B^6$ crystals, a more accurate hybrid functional was used to calculate more precise absolute valence band maximum positions. In our calculations ordinary DFT method have been used, whose limitations are well known to accurately predict excited-state properties, band gap and absolute energy positions of the valence band maximum. Nevertheless, we found that (Tab. 4) the energy difference between the valence band maxima of the various compounds of this monochalcogenides group quite close to the results obtained by a more advanced method used in [5]. We believe that this situation takes place for arbitrary thickness of monochalcogenides crystals of gallium and indium. Thus, the results of our calculations can be used to calculate the contact potential difference of these crystals. In particular, we found that the difference between the maxima of the valence bands of GaSe and InSe with large thicknesses is about 0.2 eV, which is close to the values estimated from the experiments on electron diffraction, X-ray photoelectronic spectroscopy [11], and electroluminescence [12].

## 4. Conclusion

This is first report of an ab initio local density functional calculations of the thickness dependences of the valence band offsets of β- polytype monochalcogenides GaS, GaSe, and InSe. The method of periodic slabs separated by vacuum of two-layer width was used to model a finite thickness of a crystal. It has been shown that the valence band offsets are practically independent of the crystal thickness when the latter exceeds 12 layers.

**Table 1.** Equilibrium values of the parameters of crystal structure of GaS for various thicknesses

| Number of layers | Lattice parameter (Å) $a$ | Interatomic distances (Å) $d_{Ga-Ga}$ | $d_{Ga-S}$ | Layer thickness (Å) | Crystal thickness (Å) | Interlayer distance (Å) | Vacuum width (Å) |
|---|---|---|---|---|---|---|---|
| 1* | 3.58 | 2.45 | 2.33 | | | - | 18 |
| 1 | 3.487 | 2.36 | 2.27 | 4.47 | 4.47 | - | 25.74 |
| 2 | 3.489 | 2.36 | 2.27 | 4.47 | 12.04 | 3.096 | 18.30 |
| 4 | 3.49 | 2.36 | 2.27 | 4.47 | 27.18 | 3.096 | 18.44 |
| 6 | 3.49 | 2.36 | 2.275 | 4.47 | 42.32 | 3.096 | 18.22 |
| 8 | 3.49 | 2.36 | 2.275 | 4.47 | 52.99 | 3.096 | 18.41 |
| 10 | 3.49 | 2.36 | 2.275 | 4.47 | 72.60 | 3.097 | 18.24 |
| 12 | 3.49 | 2.36 | 2.28 | 4.47 | 87.74 | 3.10 | 18.24 |

Note: 1* are taken from [5].

**Table 2.** Equilibrium values of the parameters of crystal structure of GaSe for various thicknesses

| Number of layers | Lattice parameter (Å) $a$ | Interatomic distances (Å) $d_{Ga-Ga}$ | $d_{Ga-Se}$ | Layer thickness (Å) | Crystal thickness (Å) | Interlayer distance (Å) | Vacuum width (Å) |
|---|---|---|---|---|---|---|---|
| 1* | 3.75 | 2.44 | 2.46 | | | - | 18.00 |
| 1 | 3.66 | 2.35 | 2.40 | 4.64 | 4.64 | - | 26.71 |
| 2 | 3.67 | 2.35 | 2.41 | 4.64 | 12.47 | 3.20 | 18.87 |
| 4 | 3.67 | 2.35 | 2.41 | 4.64 | 28.13 | 3.19 | 18.87 |
| 6 | 3.67 | 2.35 | 2.41 | 4.64 | 43.79 | 3.19 | 18.86 |
| 8 | 3.67 | 2.35 | 2.41 | 4.64 | 59.48 | 3.19 | 18.84 |
| 10 | 3.67 | 2.35 | 2.41 | 4.64 | 75.11 | 3.19 | 18.84 |
| 12 | 3.67 | 2.35 | 2.41 | 4.64 | 90.77 | 3.19 | 18.84 |

**Table 3.** Equilibrium values of the parameters of crystal structure of InSe for various thicknesses

| Number of layers | Lattice parameter (Å) $a$ | Interatomic distances (Å) $d_{In-In}$ | Interatomic distances (Å) $d_{In-Se}$ | Layer thickness (Å) | Crystal thickness (Å) | Interlayer distance (Å) | Vacuum width (Å) |
|---|---|---|---|---|---|---|---|
| 1* | 4.02 | 2.77 | 2.65 |  |  | - | 18 |
| 1 | 3.90 | 2.66 | 2.57 | 5.15 | 5.15 | - | 27.57 |
| 2 | 3.91 | 2.67 | 2.57 | 5.15 | 13.32 | 3.03 | 19.39 |
| 4 | 3.91 | 2.67 | 2.58 | 5.14 | 29.66 | 3.02 | 19.36 |
| 6 | 3.91 | 2.67 | 2.58 | 5.14 | 46.00 | 3.03 | 19.37 |
| 8 | 3.91 | 2.67 | 2.58 | 5.14 | 62.34 | 3.02 | 19.38 |
| 10 | 3.91 | 2.67 | 2.58 | 5.14 | 78.61 | 3.02 | 19.39 |
| 12 | 3.91 | 2.67 | 2.58 | 5.15 | 95.10 | 3.02 | 19.40 |

**Table 4.** Calculated valence band offset (VBO) for monolayers of GaS, GaSe, and InSe.

| Crystal contact | VBO (eV) | VBO[a] (eV) |
|---|---|---|
| GaSe/GaS | 0.39 | 0.41 |
| GaSe/InSe | 0.14 | 0.15 |
| InSe/GaS | 0.25 | 0.26 |

[a]Ref.[5]

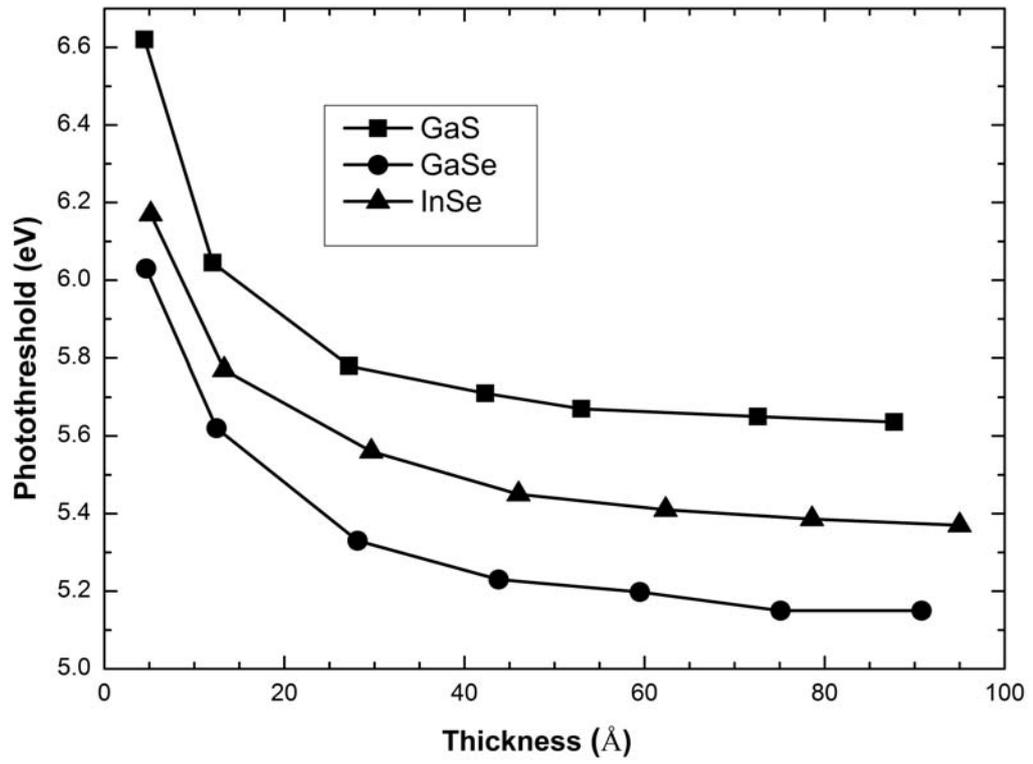

**Fig. 1.** Photothreshold as a function of the thickness of GaS, InSe, GaSe crystals.